# Modeling and Analysis of Non-Orthogonal MBMS Transmission in Heterogeneous Networks

Zhengquan Zhang, *Student Member, IEEE,* Zheng Ma, *Member, IEEE,* Ming Xiao, *Senior Member, IEEE,* and Pingzhi Fan, *Fellow, IEEE*

## Abstract

Broadcasting/multicasting is an efficient mechanism for multimedia communications due to its high spectrum efficiency, which achieves point-to-multipoint transmission on the same radio resources. To satisfy the increasing demands for multimedia broadcast multicast service (MBMS), we present a power domain non-orthogonal MBMS transmission scheme in a $K$-tier heterogeneous network (HetNet). Firstly, the system model, usage scenarios, and fundamentals of the presented scheme are discussed. Next, a tractable framework is developed to analyse the performance of non-orthogonal MBMS transmission, by using stochastic geometry. Based on this framework, the analytical expressions for the signal-to-interference-plus-noise ratio (SINR) coverage probability, average number of served users, and sum rate are derived. Furthermore, synchronous non-orthogonal MBMS transmission to further improving the system performance is also studied. The results demonstrate that non-orthogonal MBMS transmission can achieve better performance than the conventional one, in which non-orthogonal multi-rate one can fully utilize channel conditions to achieve a significant rate gain, while non-orthogonal multi-service one can efficiently use power resources to guarantee the quality of service (QoS) of high priority users, and also provide services for low priority users simultaneously.

## Index Terms

This work was supported by the National Natural Science Foundation of China under Grant 61571373, the National Science and Technology Major Project under Grant 2016ZX03001018-002, the National High-Tech R&D Program of China under Grant 2014AA01A707, and the 111 Project under Grant 111-2-14.

Z. Zhang, Z. Ma, and P. Fan are with the Key Lab of Information Coding and Transmission, Southwest Jiaotong University, Chengdu, 610031, China (e-mail: zhang.zhengquan@hotmail.com; zma@home.swjtu.edu.cn; pzfan@home.swjtu.edu.cn).

M. Xiao is with the Communication Theory Lab, School of Electrical Engineering and the ACCESS Linnaeus Center, Royal Institute of Technology, Sweden (e-mail: mingx@kth.se).

## I. INTRODUCTION

The explosive growth of mobile data traffic, especially video services, requires the development of more efficient wireless communication technologies for the year 2020 and beyond [1], [2]. Broadcasting/multicasting [3]–[7] as a point-to-multipoint (PtM) transmission, can deliver the same content to all users or a specific group of users on the same radio resources. Thus, it is an efficient mechanism for multimedia communications, and is also an important solution to satisfying the increasing demands for high data rate in future wireless networks. Due to its excellent spectrum efficiency, this technique has already been adopted by 3GPP networks as one of the key enablers to deliver multimedia services, named as multimedia broadcast multicast service (MBMS) [8]. However, conventional MBMS transmission cannot fully utilize some resource domains, e.g., power domain, due to the employment of orthogonal multiplexing. Moreover, the data rate for this orthogonal MBMS transmission (OMT) mainly relies on the weak users who have the low signal-to-interference-plus-noise ratio (SINR), in order to ensure that all users can successfully receive the same content. To address this drawback and satisfy the increasing demands for multimedia communications, the development of highly efficient MBMS transmission has attracted an increasing interest [9].

Recently, the emerging non-orthogonal multiplexing technologies [10]–[12], such as power domain non-orthogonal multiple access (NOMA) and code domain sparse code multiple access (SCMA), provides a feasible solution to improving the performance of conventional MBMS transmission. This technique multiplexes multiple users in certain resource domains, such that it can significantly improve the spectrum efficiency, reduce the transmission latency, and support massive connectivity. Especially, power domain NOMA which multiplexes multiple users in the power domain by superposition coding and decodes the desired data from the superposed signal through successive interference cancellation (SIC) [3], has recently attracted great interest and been adopted by 3GPP networks to enhance downlink transmission, named as multi-user superposition transmission (MUST) [11].

### A. Related Literature

Some recent efforts have been devoted to the performance analysis and optimization of NOMA, including sum rate [13], [14], outage probability [13]–[16], and energy efficiency [17]. In [13], the sum rate and outage probability of NOMA with randomly distributed users, were studied. Considering NOMA with partial channel state information (CSI), the authors in [14] investigated the sum rate and outage probability. The work in [15] analysed the outage probability of the downlink NOMA systems with one-bit CSI feedback. Furthermore, cooperative NOMA was proposed to improve the outage probability of the weak user [16], in which the strong user as a relay cooperates data transmission for the weak one. In addition, considering QoS constraints, the authors in [17] discussed the energy efficiency of the uplink and downlink NOMA systems, and studied optimal power division and allocation strategies. The authors of [18], [19] also studied the NOMA performance in the multiple input and multiple output (MIMO) systems. Note that different user pairing and power allocation schemes have different effects on the performance of NOMA systems. Therefore, the work in [20] studied three user pairing and power allocation schemes: fractional transmit power allocation (FTPA), pre-defined user grouping and per-group fixed power allocation (FPA), and full search power allocation (FSPA). In [21], comparison was made between NOMA with cognitive radio inspired NOMA (CR-NOMA) and fixed power allocation (F-NOMA) for downlink transmissions. Furthermore, some optimal power allocation and subchannel assignment schemes were also studied to improve the performance of NOMA systems [22], [23]. Finally, the application of NOMA to broadcasting/multicasting [24]–[26], cooperative communications [27], millimeter wave communications [28], internet of things (IoT) [29], cognitive radio networks [30], and simultaneous wireless information and power transfer (SWIPT) [31], were studied.

Especially, the application of non-orthogonal transmission to the digital broadcasting systems, has been studied by advanced television systems committee (ATSC). One of the key features of the new standard ATSC3.0 is that the digital TV systems can utilize the traditional TV channels to provide multimedia services with high data rate for mobile users simultaneously, through non-orthogonal transmission. The authors in [26] presented a comprehensive overview on non-orthogonal transmission in the digital TV ATSC 3.0 systems, named as layered division multiplexing (LDM). The ATSC3.0 systems specify two-layer LDM structure, in which the core layer is used to serve the fixed TV reception terminals, while the enhanced layer carries mobile services in one 6 MHz channel. However, compared with the digital TV systems with large coverage based on high power high tower (HPHT) [9], the wireless networks with multi-tier dense cell deployments are more complicated. Currently, there lacks the systematic study on non-orthogonal MBMS transmission (NOMT) in the wireless networks.

*B. Main Contribution*

Based on these observations, we study in detail the performance of power domain non-orthogonal MBMS transmission in a $K$-tier single-frequency heterogeneous network (HetNet) [32], [33], by using stochastic geometry [32]–[35]. The main contributions are summarized as follows.

- To begin with, we present a non-orthogonal MBMS transmission scheme in a $K$-tier single-frequency HetNet, which divides the power domain into multiple layers with different power levels and each power layer carries different MBMS content with different priority levels. We also identify two main usage scenarios: non-orthogonal multi-service MBMS transmission (NOMSMT) and non-orthogonal multi-rate MBMS transmission (NOMRMT), and discuss their fundamental principles.

- Furthermore, we develop a tractable framework to analyse the performance of non-orthogonal MBMS transmission in a $K$-tier single-frequency HetNet, by using stochastic geometry. Based on this framework, we derive the analytical expressions for the SINR coverage probability, average number of served users, and sum rate, to evaluate the performance of the presented scheme. The results demonstrate that non-orthogonal MBMS transmission can significantly improve the system performance.

- Finally, we study synchronous non-orthogonal MBMS transmission in a $K$-tier single-frequency HetNet to further improve the system performance, which enables all BSs in the MBMS service area to transmit the same content on the same radio resources, such that the signals received by the users are strengthened, instead of interfered. We also derive the analytical expressions to evaluate its performance based on the developed framework, and compare its performance with the asynchronous one. The results reveal different effects of asynchronous and synchronous transmission on the performance of non-orthogonal MBMS delivery.

*C. Paper Outline*

The rest of the paper is organized as follows. Section II describes the system model and explains the main concept of non-orthogonal MBMS transmission. The detailed performance analyses of asynchronous and synchronous non-orthogonal MBMS transmission in a $K$-tier single frequency HetNet are presented in Section III and IV, respectively. Analytical results,

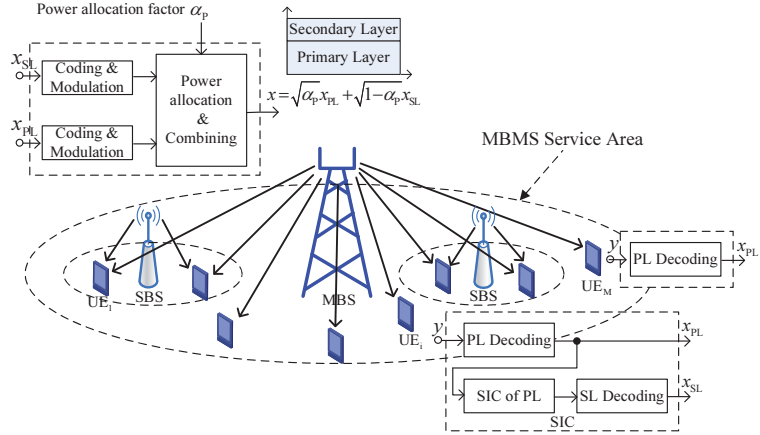

Fig. 1. System model of non-orthogonal MBMS transmission in a single frequency HetNet

Monte Carlo simulations and discussion are presented in Section V, followed by the conclusions in Section VI.

## II. NON-ORTHOGONAL MBMS TRANSMISSION

### A. System Model

Fig. 1 illustrates the system model of non-orthogonal MBMS transmission in a single frequency HetNet. With superposition coding [3], multiple layers with different power levels can be multiplexed in the power domain, and form a superposed signal. Without loss of generality, two-layer non-orthogonal MBMS transmission is assumed. The primary layer (PL) carries the high priority (HP) content with more power, while the rest of power is allocated to the secondary layer (SL) of low priority (LP). The content carried by each power layer depends on the specific usage scenarios. The basic processing of non-orthogonal MBMS transmission at the network side is: The two data streams, after the independent processing of channel coding and modulation, are combined together with power allocation and form a superposed signal. Then, this superposed signal is delivered to all interesting users. When receiving this superposed signal, for the users who only need to decode the primary layer, they directly decode it by treating the secondary layer as noise, while for the users who need to decode both the primary and secondary layers or only the secondary layer, they employ SIC to decode their desired data. To be specific, they first decode the primary layer directly by treating the secondary layer as noise, then cancel it from the received signal before decoding the secondary layer.

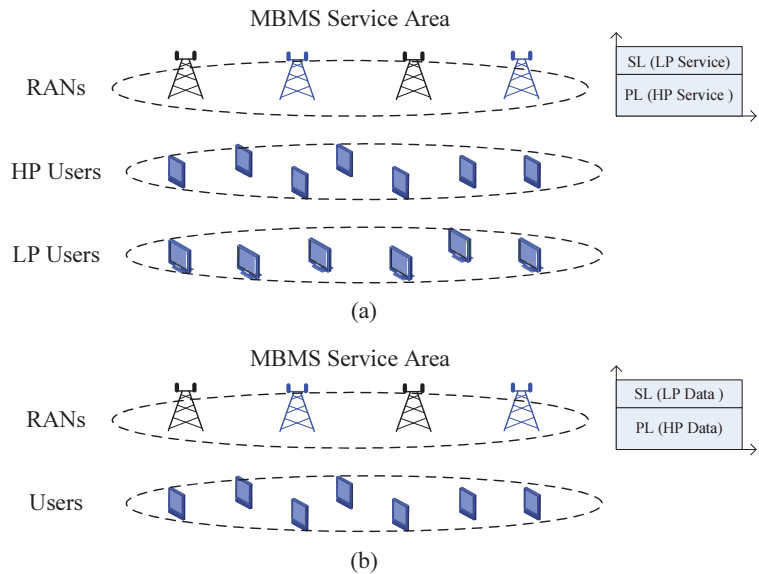

Fig. 2. Usage scenarios for non-orthogonal MBMS transmission. (a) Multi-service MBMS; (b) Multi-rate MBMS

## B. Usage Scenarios

Fig. 2 presents two main usage scenarios for non-orthogonal MBMS transmission: multi-service MBMS [26] and multi-rate MBMS [5].

*1) Multi-Service MBMS Transmission:* This scenario can be used to serve hybrid users simultaneously, e.g., mobile users and fixed TV reception terminals [26], through multiplexing a high priority service and a low priority service in the power domain. The service carried by the primary layer is delivered to the high priority users with more power, while the rest of power is allocated to the secondary layer serving the low priority users. The high priority users directly decode the primary layer to obtain the high priority service, by considering the secondary layer as noise, while the low priority users decode the secondary layer to obtain their desired service through SIC.

*2) Multi-Rate MBMS Transmission:* This scenario can guarantee the basic QoS of the weak users, and also fully utilize the channel conditions for the strong users to improve their QoS, which is very efficient for scalable multimedia transmission. Firstly, the scalable multimedia are coded into the base data and several enhanced data with different QoS requirements through source layered coding [36]. Note that enhanced data cannot work independently, without the base data. Then, these data are multiplexed orderly into different power layers to generate a superposed signal. Since the base data achieves the basic QoS, it is put in the primary layer of high priority, while the enhanced data are carried by the secondary layer of low priority. When the users receive this superposed signal, they directly decode the primary layer to obtain the base data, and then try to decode the secondary layer through SIC. More specifically, for the users with weak channel conditions, they can decode the primary layer to obtain the basic QoS, while the users with strong channel conditions can further decode the secondary layer to obtain the enhanced data, such that they can obtain enhanced QoS. Note that for non-scalable multimedia services with non-orthogonal transmission, the BSs distribute the identical service to all users with low and high data rates simultaneously, through different power layers, such that the weak users decode the primary layer with low data rate to obtain the basic QoS, while the strong users can decode the secondary layer with high data rate through SIC, to obtain better QoS.

### C. Fundamentals of Non-Orthogonal Transmission

We assume that $R_{PL}$ and $R_{SL}$ represent the data transmission rates for the primary and secondary layers, respectively. At the network side, the primary and secondary layers carry multiple service data and are multiplexed in the power domain as a superposed signal, which can be expressed as

$$x = \sqrt{\alpha_p} x_{PL} + \sqrt{1 - \alpha_p} x_{SL}, \quad (1)$$

where $0 < \alpha_p < 1$ is power allocation factor (PAF), $x_{PL}$ and $x_{SL}$ are signals of the primary and secondary layers, respectively. Note that when $\alpha_p = 1$, power domain NOMA degrades to orthogonal multiple access (OMA).

In the $k$-th time slot, the user, $\text{UE}_i$, listens to the interesting services on the given radio resources. Thus, the signal received at $\text{UE}_i$ with a random distance, $d$, can be expressed as

$$y_i[k] = h_i \sqrt{P} d^{-\alpha/2} x[k] + n_i[k] = h_i \sqrt{P} d^{-\alpha/2} (\sqrt{\alpha_p} x_{PL} + \sqrt{1 - \alpha_p} x_{SL}) + n_i[k], \quad (2)$$

where $h$ is the small-scale fading coefficient with the distribution, $|h|^2 \sim \exp(1)$, $P$ is the transmit power, $d$ is the distance between the user and its serving BS, $\alpha > 2$ is the path loss exponent, and $n \sim CN(0, \sigma^2)$ is the additive white Gaussian noise (AWGN).

SIC is employed to decode the desired data from the superposed signal: The user first decodes the primary layer by treating the secondary layer as noise, then cancels it from the received signal before further decoding the secondary layer. Therefore, the instantaneous SINRs of detecting the primary and secondary layers can be expressed, respectively, as

$$\gamma_i^{PL} = \frac{\alpha_p |h_i|^2 d^{-\alpha} P}{(1 - \alpha_p) |h_i|^2 d^{-\alpha} P + \sigma^2}, \quad (3)$$

and
$$\gamma_i^{SL} = \frac{(1-\alpha_p)|h_i|^2 d^{-\alpha} P}{\sigma^2}. \tag{4}$$

*1) Sum Rate:* The user, $UE_i$, decodes the data successfully, if the instantaneous channel capacity is larger than the rate threshold. Therefore, the achievable data rates for decoding the primary and secondary layers can be written as

$$R_i^{PL} = \begin{cases} R_{PL}, & c(\gamma_i^{PL}) \geq R_{PL}, \\ 0, & c(\gamma_i^{PL}) < R_{PL}, \end{cases} \tag{5}$$

and
$$R_i^{SL} = \begin{cases} R_{SL}, & c(\gamma_i^{SL}) \geq R_{SL} \ \&\& \ c(\gamma_i^{PL}) \geq R_{PL}, \\ 0, & c(\gamma_i^{SL}) < R_{SL} \ || \ c(\gamma_i^{PL}) < R_{PL}, \end{cases} \tag{6}$$

where $c(x) = \log_2(1+x)$ is the channel capacity.

For non-orthogonal multi-service MBMS transmission, the high priority users are served by the primary layer with high priority, while the secondary layer is used to serve the low priority users. Thus, the sum rate for certain MBMS area is defined as the total rate for all high and low priority users in that area, and can be expressed as

$$R_{sum} = \underbrace{\sum_i R_i^{PL}}_{\text{High priority users}} + \underbrace{\sum_j R_j^{SL}}_{\text{Low priority users}}, \tag{7}$$

while the sum rate for non-orthogonal multi-rate MBMS transmission is

$$R_{sum} = \sum_i (R_i^{PL} + R_i^{SL}). \tag{8}$$

*2) Outage Probability:* The maximum SINR, that the user, $UE_i$, detects the primary layer, is $\lim_{h \to \infty} \gamma_i^{PL} = \frac{\alpha_p}{1-\alpha_p}$. Therefore, the probability that $UE_i$ cannot decode the primary layer is

$$P_{out,i}^{PL} = Pr\{\log_2(1+\gamma_i^{PL}) < R_{PL}\} = \begin{cases} 1, & T_{PL} \geq \frac{\alpha_p}{1-\alpha_p}, \\ 1 - e^{-\frac{T_{PL} d^\alpha \sigma^2}{P(\alpha_p - T_{PL}(1-\alpha_p))}}, & 0 < T_{PL} < \frac{\alpha_p}{1-\alpha_p}, \end{cases} \tag{9}$$

where $T_{PL} = 2^{R_{PL}} - 1$ is the SINR threshold for the primary layer. According to SIC, the probability that $UE_i$ cannot decode the secondary layer is

$$\begin{aligned} P_{out,i}^{SL} &= Pr\{(\gamma_i^{PL} < T_{PL}) \ || \ (\gamma_i^{SL} < T_{SL})\} \\ &= \begin{cases} 1, & T_{PL} \geq \frac{\alpha_p}{1-\alpha_p}, \\ 1 - e^{-\frac{T_{PL} d^\alpha \sigma^2}{P(\alpha_p - T_{PL}(1-\alpha_p))}}, & 0 < T_{PL} < \frac{\alpha_p}{1-\alpha_p} \ \&\& \ T_{SL} \leq \frac{(1-\alpha_p)T_{PL}}{\alpha_p - (1-\alpha_p)T_{PL}}, \\ 1 - e^{-\frac{T_{SL} d^\alpha \sigma^2}{(1-\alpha_p)P}}, & 0 < T_{PL} < \frac{\alpha_p}{1-\alpha_p} \ \&\& \ T_{SL} > \frac{(1-\alpha_p)T_{PL}}{\alpha_p - (1-\alpha_p)T_{PL}}, \end{cases} \end{aligned} \tag{10}$$

where $T_{SL} = 2^{R_{SL}} - 1$ is the SINR threshold for the secondary layer.

For multi-service MBMS transmission, the high priority users decode the primary layer, while the low priority users decode the secondary layer. Thus, the corresponding outage probabilities of the high and low priority users, $\text{UE}_{HP,i}$ and and $\text{UE}_{LP,i}$, can be expressed as

$$P_{out,HP,i} = P_{out,i}^{PL} = \begin{cases} 1, \ T_{PL} \geq \frac{\alpha_p}{1-\alpha_p}, \\ 1 - e^{-\frac{T_{PL} d^\alpha \sigma^2}{P(\alpha_p - T_{PL}(1-\alpha_p))}}, \ 0 < T_{PL} < \frac{\alpha_p}{1-\alpha_p}, \end{cases} \quad (11)$$

and

$$P_{out,LP,i} = P_{out,i}^{SL} = \begin{cases} 1, \ T_{PL} \geq \frac{\alpha_p}{1-\alpha_p}, \\ 1 - e^{-\frac{T_{PL} d^\alpha \sigma^2}{P(\alpha_p - T_{PL}(1-\alpha_p))}}, \ 0 < T_{PL} < \frac{\alpha_p}{1-\alpha_p} \ \&\& \ T_{SL} \leq \frac{(1-\alpha_p)T_{PL}}{\alpha_p - (1-\alpha_p)T_{PL}}, \\ 1 - e^{-\frac{T_{SL} d^\alpha \sigma^2}{(1-\alpha_p)P}}, \ 0 < T_{PL} < \frac{\alpha_p}{1-\alpha_p} \ \&\& \ T_{SL} > \frac{(1-\alpha_p)T_{PL}}{\alpha_p - (1-\alpha_p)T_{PL}}. \end{cases} \quad (12)$$

Note that the path loss exponents for the high and low priority users may be different. In that case, we can obtain the expressions for the results, by replacing the path loss exponent, $\alpha$, in (11) by that of high priority users, and substituting that of low priority users for $\alpha$ in (12).

For multi-rate MBMS transmission, the users first decode the primary layer to obtain the basic QoS, and further decode the secondary layer to obtain the enhanced data to improve QoS. Note that the users do not decode the secondary layer, if they have failed to decode the primary layer. This is because the enhanced data cannot work independently without the base data. Therefore, if the users can decode the primary layer, it can transfer information. Accordingly, the outage probability of multi-rate MBMS transmission can be expressed as

$$P_{out,i} = P_{out,i}^{PL} = \begin{cases} 1, \ T_{PL} \geq \frac{\alpha_p}{1-\alpha_p}, \\ 1 - e^{-\frac{T_{PL} d^\alpha \sigma^2}{P(\alpha_p - T_{PL}(1-\alpha_p))}}, \ 0 < T_{PL} < \frac{\alpha_p}{1-\alpha_p}. \end{cases} \quad (13)$$

### III. NON-ORTHOGONAL MBMS TRANSMISSION IN HETNETS

*A. Non-Orthogonal Multi-Rate MBMS Transmission*

*1) Network Model:* The important parameters related to the model of non-orthogonal MBMS transmission in a $K$-tier HetNet are shown as follows. Similar to [33]–[35], the BS locations of the $k$-th network tier follow an independently homogeneous PPP, $\Phi_{B_k}$, with density, $\lambda_{B_k}$. Note that in general, we have $\lambda_{B_1} < \lambda_{B_2} < ... < \lambda_{B_K}$. The user locations are also modeled as an independently homogeneous PPP, $\Phi_U$, with density, $\lambda_U$. The transmit power of the BSs in the $k$-th network tier is assumed to be $P_{B_k}$ and $P_{B_1} > ... > P_{B_K}$, while $\sigma^2$ is the noise power. According to the standard power loss propagation model [34], the average power received at a user with certain distance, $\|X\|$, from one BS of the $k$-th network tier, is $P_{\text{Rx}}(\|X\|) = P_{B_k}\|X\|^{-\alpha}$, where $\alpha > 2$ is the path loss exponent. The small-scale fading is subject to Rayleigh fading, with the distribution, $H = |h|^2 \sim \exp(1)$.

For MBMS transmission, the users are associated with one BS in a $K$-tier HetNet based on the maximum average receiving power (MARP) [32] without bias. Thus, the probability that the user is associated with the $k$-th network tier can be expressed as [32]

$$\mathcal{A}_k = 2\pi\lambda_{B_k} \int_0^\infty r \exp\left(-\pi \sum_{i=1}^K \lambda_{B_i} \hat{P}_{B_i}^{2/\alpha} r^2\right) dr. \tag{14}$$

The user with a random distance, $\|X\|$, receives not only the desired signal from its serving BS, $B_o$, located at the $k$-th network tier, but also the co-channel interference (CCI) from other BSs. Therefore, the sum signal received at the user can be expressed as

$$y_k = h_k P_{B_k}^{1/2} \|X\|^{-\alpha/2} x + \underbrace{\sum_{k=1}^K \sum_{X_{k,j} \in \Phi_{B_k} \setminus B_o} h_{k,j} P_{B_k}^{1/2} \|X_{k,j}\|^{-\alpha/2} x_{k,j}}_{I_{CCI}} + n, \tag{15}$$

where $X_{k,j}$ is the location of the $j$-th neighboring BS in the $k$-th network tier. Note that for a $K$-tier single-frequency HetNet, the user suffers the co-channel interference not only from intra-tier BSs, but also from the inter-tier BSs. If different carrier frequencies are used in different network tiers, there is no inter-tier interference.

Using SIC, the SINRs that the user detects the primary and secondary layers provided by the $k$-th network tier can be expressed as

$$SINR_{PL,k}^1 = \frac{\alpha_{p,k} H_k P_{B_k} \|X\|^{-\alpha}}{(1-\alpha_{p,k}) H_k P_{B_k} \|X\|^{-\alpha} + I_{CCI} + \sigma^2}, \tag{16}$$

and

$$SINR_{SL,k}^1 = \frac{(1-\alpha_{p,k}) H_k P_{B_k} \|X\|^{-\alpha}}{I_{CCI} + \sigma^2}, \tag{17}$$

where, $I_{CCI} = \sum_{n=1}^K I_{CCI,n}$ is the total CCI, and $I_{CCI,n} = \sum_{X_{n,j} \in \Phi_{B_n} \setminus B_o} H_{n,j} P_{B_n} \|X_{n,j}\|^{-\alpha}$ is the CCI from the $n$-th network tier.

*2) SINR Coverage Probability:* For a $K$-tier HetNet, the user is in coverage, if the received SINRs from at least one BS of all $K$ network tiers are larger than the threshold, $T$. According to the law of total probability, the SINR coverage probability can be expressed as

$$P_c^1(T) \triangleq \sum_{k=1}^K \mathcal{A}_k P_{c,k}^1(T), \tag{18}$$

where $\mathcal{A}_k$ is the association probability to the $k$-th network tier, and $P_{c,k}^1(T)$ is the corresponding SINR coverage probability, which will be explicitly given in the following sections.

**Theorem 1.** *With fixed SINR thresholds for the primary and secondary layers, $T_{PL}$ and $T_{SL}$, the SINR coverage probability of the primary layer of NOMRMT is*

$$P_{c,PL}^1 = \sum_{k=1}^{K} \mathcal{A}_k P_{c,PL,k}^1, \qquad (19)$$

*where that of the $k$-th network tier is*

$$P_{c,PL,k}^1 = \begin{cases} 0, \ T_{PL} \geq \frac{\alpha_{p,k}}{1-\alpha_{p,k}}, \\ \frac{2\pi\lambda_{B_k}}{\mathcal{A}_k} \int_0^\infty r \exp\left(-T_{PL} C_k^{-1} SNR_k^{-1} r^\alpha \right. \\ \left. -\pi \sum_{i=1}^{K} \lambda_i \hat{P}_{B_i}^{2/\alpha}(1 + C_k^{-2/\alpha}\mathcal{Z}(T_{PL},\alpha,C_k))r^2\right)dr, \ T_{PL} < \frac{\alpha_{p,k}}{1-\alpha_{p,k}}. \end{cases} \qquad (20)$$

*with $C_k = \alpha_{p,k} - (1-\alpha_{p,k})T_{PL}$, $SNR_k = P_{B_k}/\sigma^2$, and $\mathcal{Z}(T,\alpha,C_k) = T^{2/\alpha} \int_{(C_k/T)^{2/\alpha}}^\infty \frac{1}{1+t^{\alpha/2}}dt$.*

*The SINR coverage probability of the secondary layer of NOMRMT is*

$$P_{c,PSL}^1 = \sum_{k=1}^{K} \mathcal{A}_k P_{c,PSL,k}^1, \qquad (21)$$

*where that of the $k$-th network tier is*

$$P_{c,PSL,k}^1 = \begin{cases} 0, \ T_{PL} \geq \frac{\alpha_{p,k}}{1-\alpha_{p,k}}, \\ \frac{2\pi\lambda_{B_k}}{\mathcal{A}_k} \int_0^\infty r \exp\left(-T_{PL} C_k^{-1} SNR_k^{-1} r^\alpha \right. \\ \left. -\pi \sum_{i=1}^{K} \lambda_i \hat{P}_{B_i}^{2/\alpha}(1 + C_k^{-2/\alpha}\mathcal{Z}(T_{PL},\alpha,C_k))r^2\right)dr, \\ \qquad T_{PL} < \frac{\alpha_{p,k}}{1-\alpha_{p,k}} \ \&\& \ T_{SL} \leq \frac{(1-\alpha_{p,k})T_{PL}}{C_k}, \\ \frac{2\pi\lambda_{B_k}}{\mathcal{A}_k} \int_0^\infty r \exp\left(-T_{SL}(1-\alpha_{p,k})^{-1} SNR_k^{-1} r^\alpha \right. \\ \left. -\pi \sum_{i=1}^{K} \lambda_{B_i} \hat{P}_{B_i}^{2/\alpha}(1 + (1-\alpha_{p,k})^{-2/\alpha}\mathcal{Z}(T_{SL},\alpha,1-\alpha_{p,k}))r^2\right)dr, \\ \qquad T_{PL} < \frac{\alpha_{p,k}}{1-\alpha_{p,k}} \ \&\& \ T_{SL} > \frac{(1-\alpha_{p,k})T_{PL}}{C_k}. \end{cases} \qquad (22)$$

*Proof.* See Appendix A. □

*3) Average Number of Served Users:* The analysis is for a typical MBMS area, $\mathcal{A}(0, R_{TA})$, at the origin with radius, $R_{TA} = (\pi\lambda_B)^{-1/2}$. In general, $\lambda_B$ is equal to the macro BS density, i.e., $\lambda_{B_1}$. For a $K$-tier HetNet, the users in the typical MBMS area can be served by each network tier. According to the law of total probability, the average number of served users in the typical MBMS area can be expressed as

$$\mathbb{E}^o[N^1] \triangleq \sum_{k=1}^{K} \mathcal{A}_k \mathbb{E}^o[N_k^1], \qquad (23)$$

where $\mathcal{A}_k$ is the probability associated with the $k$-th network tier, and $\mathbb{E}^o[N_k^1]$ is the corresponding average number of served users

$$\mathbb{E}^o[N_k^1] \triangleq \mathbb{E}^o\left[\sum_{y_{k,j}\in\Phi_{U_k,\mathcal{A}(0,R_{TA})}} \mathbb{I}(P_{c,y_{k,j}}(T))\right]. \tag{24}$$

**Proposition 1.** *The average number of served users by the primary and secondary layers of NOMRMT in the typical MBMS area are*

$$\mathbb{E}^o[N_{PL}^1] = \lambda_B^{-1}\lambda_U \sum_{k=1}^K \mathcal{A}_k P_{c,PL,k}^1, \tag{25}$$

*and*

$$\mathbb{E}^o[N_{PSL}^1] = \lambda_B^{-1}\lambda_U \sum_{k=1}^K \mathcal{A}_k P_{c,PSL,k}^1. \tag{26}$$

*Proof.* See Appendix B. □

*4) Sum Rate:* The sum rate for the $k$-th network tier of the typical MBMS area is the mean of total rate for all users associated to that tier

$$R_{sum,k}^1 = R_{PL}\mathbb{E}^o[N_{PL,k}^1] + R_{SL}\mathbb{E}^o[N_{PSL,k}^1], \tag{27}$$

while the sum rate for the typical MBMS area is

$$R_{sum}^1 = R_{PL}\sum_{k=1}^K \mathbb{E}^o[N_{PL,k}^1] + R_{SL}\sum_{k=1}^K \mathbb{E}^o[N_{PSL,k}^1]. \tag{28}$$

**Proposition 2.** *With fixed rate thresholds for the primary and secondary layers, $R_{PL}$ and $R_{SL}$, the sum rate for NOMRMT in the typical MBMS area is*

$$R_{sum}^1 = R_{PL}\lambda_B^{-1}\lambda_U \sum_{k=1}^K \mathcal{A}_k P_{c,PL,k}^1 + R_{SL}\lambda_B^{-1}\lambda_U \sum_{k=1}^K \mathcal{A}_k P_{c,PSL,k}^1. \tag{29}$$

*Proof.* Combining (28), (72), and (73), the sum rate can be obtained as in (29) and the proof is completed. □

### B. Non-Orthogonal Multi-Service MBMS Transmission

*1) Network Model:* Without loss of generality, we consider two user tiers in a $K$-tier HetNet, where the first user tier is high priority users, while the second user tier is low priority users. Taking mobile users and fixed TV users for example, we assume that mobile users have high priority, while low priority is for fixed TV users. Therefore, the primary layer carries the content for mobile users, while the secondary layer is used to serve the fixed TV users. The important parameters related to this network model are shown as follows. The $k$-th MBS locations comply with an independently homogeneous PPP, $\Phi_{B_k}$, with density, $\lambda_{B_k}$, while the user locations of the $i$-th user tier also follow an independently homogeneous PPP, $\Phi_{U_i}$, with density, $\lambda_{U_i}$, respectively. The path loss exponents from the BSs to the high and low priority users are assumed to be, $\alpha_1$, and $\alpha_2$, respectively. Furthermore, the small-scale fading is subject to Rayleigh fading.

The SINRs of the primary and secondary layers received at the user in the $i$-th user tier served by the $k$-th network tier can be expressed, respectively, as

$$SINR_{PL,k,i}^2 = \frac{\alpha_{p,k} H_k P_{B_k} \|X\|^{-\alpha_i}}{(1-\alpha_{p,k}) H_k P_{B_k} \|X\|^{-\alpha_i} + I_{CCI,i} + \sigma^2}, \tag{30}$$

and

$$SINR_{SL,k,i}^2 = \frac{(1-\alpha_{p,k}) H_k P_{B_k} \|X\|^{-\alpha_i}}{I_{CCI,i} + \sigma^2}, \tag{31}$$

where $I_{CCI,i} = \sum_{j=1}^{K} I_{CCI,i,j}$ is the total CCI, and $I_{CCI,i,k} = \sum_{X_{k,j} \in \Phi_{B_k} \setminus B_o} H_{k,j} P_{B_k} \|X_{k,j}\|^{-\alpha_i}$ is the CCI from the $k$-th network tier.

*2) SINR Coverage Probability:*

**Theorem 2.** *With fixed SINR thresholds for the high and low priority users, $T_{HP}$ and $T_{LP}$, the SINR coverage probability of the high priority users is*

$$P_{c,HP}^2(T_{HP}) = \sum_{k=1}^{K} \mathcal{A}_k P_{c,HP,k}^2, \tag{32}$$

*where that of the $k$-th network tier is*

$$P_{c,HP,k}^2 = \begin{cases} 0, \; T_{HP} \geq \frac{\alpha_{p,k}}{1-\alpha_{p,k}}, \\ \frac{2\pi \lambda_{B_k}}{\mathcal{A}_k} \int_0^\infty r \exp\left(-T_{HP} C_k^{-1} SNR_k^{-1} r^{\alpha_1} \right. \\ \left. -\pi \sum_{i=1}^{K} \lambda_i \hat{P}_{B_i}^{2/\alpha_1} (1 + C_k^{-2/\alpha_1} \mathcal{Z}(T_{HP}, \alpha_1, C_k)) r^2 \right) dr, \; T_{HP} < \frac{\alpha_{p,k}}{1-\alpha_{p,k}}. \end{cases} \tag{33}$$

*where $C_k = \alpha_{p,k} - (1-\alpha_{p,k}) T_{HP}$, and $SNR_k = P_{B_k}/\sigma^2$, and $\mathcal{Z}(T, \alpha, C_k) = T^{2/\alpha} \int_{(C_k/T)^{2/\alpha}}^{\infty} \frac{1}{1+t^{\alpha/2}} dt$.*

*The SINR coverage probability of the low priority users is*

$$P_{c,LP}^2(T_{LP}) = \sum_{k=1}^{K} \mathcal{A}_k P_{c,LP,k}^2, \tag{34}$$

*where that of the $k$-th network tier is*

$$P_{c,LP,k}^2 = \begin{cases} 0, \ T_{HP} \geq \frac{\alpha_{p,k}}{1-\alpha_{p,k}}, \\ \frac{2\pi\lambda_{B_k}}{\mathcal{A}_k} \int_0^\infty r \exp\left(-T_{HP}C_k^{-1}SNR_k^{-1}r^{\alpha_2} \right. \\ \left. -\pi \sum\limits_{i=1}^{K} \lambda_i \hat{P}_{B_i}^{2/\alpha_2}(1+C_k^{-2/\alpha_2}\mathcal{Z}(T_{PL},\alpha_2,C_k))r^2\right) dr, \\ \qquad T_{HP} < \frac{\alpha_{p,k}}{1-\alpha_{p,k}} \ \&\& \ T_{LP} \leq \frac{(1-\alpha_{p,k})T_{HP}}{C_k}, \\ \frac{2\pi\lambda_{B_k}}{\mathcal{A}_k} \int_0^\infty r \exp\left(-T_{LP}(1-\alpha_{p,k})^{-1}SNR_k^{-1}r^{\alpha_2} \right. \\ \left. -\pi \sum\limits_{i=1}^{K} \lambda_{B_i} \hat{P}_{B_i}^{2/\alpha_2}(1+(1-\alpha_{p,k})^{-2/\alpha_2}\mathcal{Z}(T_{LP},\alpha_2,1-\alpha_{p,k}))r^2\right) dr, \\ \qquad T_{HP} < \frac{\alpha_{p,k}}{1-\alpha_{p,k}} \ \&\& \ T_{LP} > \frac{(1-\alpha_{p,k})T_{HP}}{C_k}. \end{cases} \quad (35)$$

*Proof.* Similar to Theorem 1, replacing $\alpha$ in (20) with $\alpha_1$, (33) can be obtained. The path loss exponent, $\alpha$, in (22) is replaced by $\alpha_2$, to obtain (35) and the proof is completed. □

*3) Average Number of Served Users:*

**Proposition 3.** *The average number of served high and low priority users of NOMSMT in the typical MBMS area can be expressed, respectively, as*

$$\mathbb{E}^o[N_{HP}^2] = \lambda_B^{-1}\lambda_{U_1} \sum_{k=1}^{K} \mathcal{A}_k P_{c,HP,k}^2, \quad (36)$$

*and*

$$\mathbb{E}^o[N_{LP}^2] = \lambda_B^{-1}\lambda_{U_2} \sum_{k=1}^{K} \mathcal{A}_k P_{c,LP,k}^2. \quad (37)$$

*Proof.* Similar to Proposition 1, replacing $\lambda_U$ and $P_{c,PL,k}^1$ in (25) by $\lambda_{U_1}$ and $P_{c,HP,k}^2$, the average number of served high priority users in the typical MBMS area can be obtained as in (36). Similarly, the average number of served low priority users in the typical MBMS area can be obtained as in (37), by replacing $\lambda_U$ and $P_{c,PSL,k}^1$ in (26) by $\lambda_{U_2}$ and $P_{c,LP,k}^2$. The proof is completed. □

*4) Sum Rate:* The sum rate for NOMSMT in a $K$-tier HetNet is defined as the mean of the total rate for all high and low priority user in the typical MBMS area, and can be expressed as

$$R_{sum}^2 = R_{HP} \sum_{k=1}^{K} \mathbb{E}^o[N_{HP,k}^2] + R_{LP} \sum_{k=1}^{K} \mathbb{E}^o[N_{LP,k}^2]. \quad (38)$$

**Proposition 4.** *With fixed rates for the high and low priority users, $R_{HP}$ and $R_{LP}$, the sum rate for NOMSMT in the typical MBMS service area is given by*

$$R_{sum}^2 = R_{HP}\lambda_B^{-1}\lambda_{U_1}\sum_{k=1}^{K}\mathcal{A}_k P_{c,HP,k}^2 + R_{LP}\lambda_B^{-1}\lambda_{U_2}\sum_{k=1}^{K}\mathcal{A}_k P_{c,LP,k}^2. \tag{39}$$

*Proof.* Combining (36), (37), and (38), the sum rate can be obtained as (39) and the proof is completed. □

## IV. SYNCHRONOUS NON-ORTHOGONAL MBMS TRANSMISSION IN HETNETS

The synchronous MBMS transmission [8] can further improve the system performance, which enables all BSs to transmit the same content on the same radio resources, such that these BSs serve the users, instead of contributing the aggregate interference. In this section, we will study the performance of synchronous non-orthogonal MBMS transmission in a $K$-tier HetNet.

### A. Non-Orthogonal Multi-Rate MBMS Transmission

*1) Network Model:* The important parameters related to the model of synchronous non-orthogonal multi-rate MBMS transmission in a $K$-tier HetNet are the same as the parameters shown in Subsection III-A1. Due to synchronous MBMS transmission, the sum signal received at the user can be expressed as

$$y = \sum_{k=1}^{K}\sum_{X_{k,j}\in\Phi_{B_k}} h_{k,j}P_{B_k}^{1/2}\|X_{k,j}\|^{-\alpha/2}x + n. \tag{40}$$

Therefore, using SIC, the corresponding SINRs of the primary and secondary layers received at the user are given by

$$SINR_{PL}^3 = \frac{\left|\sum_{k=1}^{K}\sum_{X_{k,j}\in\Phi_{B_k}}\alpha_p^{1/2}P_{B_k}^{1/2}h_{k,j}r_{k,j}^{-\alpha/2}\right|^2}{\left|\sum_{k=1}^{K}\sum_{X_{k,j}\in\Phi_{B_k}}(1-\alpha_p)^{1/2}P_{B_k}^{1/2}h_{k,j}r_{k,j}^{-\alpha/2}\right|^2 + \sigma^2}, \tag{41}$$

and

$$SINR_{SL}^3 = \frac{\left|\sum_{k=1}^{K}\sum_{X_{k,j}\in\Phi_{B_k}}(1-\alpha_p)^{1/2}P_{B_k}^{1/2}h_{k,j}r_{k,j}^{-\alpha/2}\right|^2}{\sigma^2}, \tag{42}$$

where $r_{k,j} = \|X_{k,j}\|$ is the distance between the user and the $j$-th BS in the $k$-th network tier.

*2) SINR Coverage Probability:*

**Theorem 3.** *With fixed SINR thresholds for the primary and secondary layers, $T_{PL}$ and $T_{SL}$, the SINR coverage probability of the primary layer of synchronous NOMRMT can be expressed as*

$$P_{c,PL}^3 = \sum_{k=1}^{K} \mathcal{A}_k P_{c,PL,k}^3, \tag{43}$$

*where $P_{c,PL,k}^3$ is lower bounded as*

$$P_{c,PL,k}^3 \geq \begin{cases} 0, \ T_{PL} \geq \frac{\alpha_p}{1-\alpha_p}, \\ \frac{2\pi\lambda_k}{\mathcal{A}_k} \int_{r>0} r \exp\left(-\pi \sum_{n=1}^{K} \lambda_n \hat{P}_n^{2/\alpha} r^2\right) \\ \times \exp\left(-\frac{(\alpha-1)\sigma^2 T_{PL} r^{\alpha-1}}{(\alpha_p-(1-\alpha_p)T_{PL})((\alpha-1)P_{B_k} + \sum\limits_{n=1}^{K} \lambda_{B_n} P_{B_n} \hat{P}_{B_n}^{(1-\alpha)/\alpha} r)}\right) dr, \ T_{PL} < \frac{\alpha_p}{1-\alpha_p}. \end{cases} \tag{44}$$

*The SINR coverage probability of the secondary layer of synchronous NOMRMT can be expressed as*

$$P_{c,PSL}^3 = \sum_{k=1}^{K} \mathcal{A}_k P_{c,PSL,k}^3, \tag{45}$$

*where*

$$P_{c,PSL,k}^3 = \begin{cases} 0, \ T_{PL} \geq \frac{\alpha_p}{1-\alpha_p}, \\ P_{c,PL,k}^3, \ T_{PL} < \frac{\alpha_p}{1-\alpha_p} \ \&\& \ T_{SL} \leq \frac{(1-\alpha_p)T_{PL}}{C}, \\ P_{c,SL,k}^3, \ T_{PL} < \frac{\alpha_p}{1-\alpha_p} \ \&\& \ T_{SL} > \frac{(1-\alpha_p)T_{PL}}{C}, \end{cases} \tag{46}$$

*and*

$$P_{c,SL,k}^3 \geq \frac{2\pi\lambda_k}{\mathcal{A}_k} \int_{r>0} r \exp\left(-\pi \sum_{n=1}^{K} \lambda_n \hat{P}_n^{2/\alpha} r^2\right) \\ \times \exp\left(-\frac{(\alpha-1)\sigma^2 T_{PL} r^{\alpha-1}}{(1-\alpha_p)((\alpha-1)P_{B_k} + \sum\limits_{n=1}^{K} \lambda_{B_n} P_{B_n} \hat{P}_{B_n}^{(1-\alpha)/\alpha} r)}\right) dr. \tag{47}$$

*Proof.* See Appendix C □

*3) Average Number of Served Users:*

**Proposition 5.** *The average number of served users by the primary and secondary layers of synchronous NOMRMT in the typical MBMS area are*

$$\mathbb{E}^o[N_{PL}^3] = \lambda_B^{-1} \lambda_U \sum_{k=1}^{K} \mathcal{A}_k P_{c,PL,k}^3, \tag{48}$$

*and*

$$\mathbb{E}^o[N_{PSL}^3] = \lambda_B^{-1}\lambda_U \sum_{k=1}^{K} \mathcal{A}_k P_{c,PSL,k}^3. \quad (49)$$

*Proof.* Comparing the asynchronous and synchronous NOMRMT, the difference is the SINR coverage probability. Thus, the proof is similar to the asynchronous one, shown in Proposition 1. Replacing $P_{c,PL,k}^1$ and $P_{c,PSL,k}^1$ in (25) and (26) by $P_{c,PL,k}^3$ and $P_{c,PSL,k}^3$, the average number of served users can be obtained as in (48) and (49), respectively. This completes the proof. □

*4) Sum Rate:*

**Proposition 6.** *With fixed rate thresholds for the primary and secondary layers, $R_{PL}$ and $R_{SL}$, the sum rate for synchronous NOMRMT in the typical MBMS area is*

$$R_{sum}^3 = R_{PL}\lambda_B^{-1}\lambda_U \sum_{k=1}^{K} \mathcal{A}_k P_{c,PL,k}^3 + R_{SL}\lambda_B^{-1}\lambda_U \sum_{k=1}^{K} \mathcal{A}_k P_{c,PSL,k}^3. \quad (50)$$

*Proof.* Similar to the asynchronous one, shown in Proposition 2, replacing $P_{c,PL,k}^1$ and $P_{c,PSL,k}^1$ in (29) by $P_{c,PL,k}^3$ and $P_{c,PSL,k}^3$, the sum rate can be obtained as in (50) and the proof is completed. □

### B. Non-Orthogonal Multi-Service MBMS Transmission

*1) Network Model:* The important parameters related to the model of synchronous non-orthogonal multi-service MBMS transmission in a $K$-tier HetNet are the same as the parameters shown in Subsection III-B1. The SINRs received at the high and low priority users can be written, respectively, as

$$SINR_{HP}^4 = \frac{\left|\sum_{k=1}^{K}\sum_{X_{k,j}\in\Phi_{B_k}} \alpha_p^{1/2} P_{B_k}^{1/2} h_{k,j} r_{k,j}^{-\alpha_1/2}\right|^2}{\left|\sum_{k=1}^{K}\sum_{X_{k,j}\in\Phi_{B_k}} (1-\alpha_p)^{1/2} P_{B_k}^{1/2} h_{k,j} r_{k,j}^{-\alpha_1/2}\right|^2 + \sigma^2}, \quad (51)$$

*and*

$$SINR_{LP}^4 = \frac{\left|\sum_{k=1}^{K}\sum_{X_{k,j}\in\Phi_{B_k}} (1-\alpha_p)^{1/2} P_{B_k}^{1/2} h_{k,j} r_{k,j}^{-\alpha_2/2}\right|^2}{\sigma^2}. \quad (52)$$

*2) SINR Coverage Probability:*

**Theorem 4.** *With fixed SINR thresholds for the high and low priority users, $T_{HP}$ and $T_{LP}$, the SINR coverage probability of the high priority users of synchronous NOMSMT can be expressed as*

$$P_{c,HP}^4 = \sum_{k=1}^{K} \mathcal{A}_k P_{c,HP,k}^4, \tag{53}$$

*where $P_{c,HP,k}^4$ is lower bounded as*

$$P_{c,HP,k}^4 \geq \begin{cases} 0, \ T_{HP} \geq \frac{\alpha_p}{1-\alpha_p}, \\ \frac{2\pi\lambda_k}{\mathcal{A}_k} \int_{r>0} r \exp\left(-\pi \sum_{n=1}^{K} \lambda_n \hat{P}_n^{2/\alpha_1} r^2\right) \\ \times \exp\left(-\frac{(\alpha_1-1)\sigma^2 T_{HP} r^{\alpha_1-1}}{(\alpha_p-(1-\alpha_p)T_{HP})((\alpha_1-1)P_{B_k}+\sum_{n=1}^{K} \lambda_{B_n} P_{B_n} \hat{P}_{B_n}^{(1-\alpha_1)/\alpha_1} r)}\right) dr, \ T_{HP} < \frac{\alpha_p}{1-\alpha_p}. \end{cases} \tag{54}$$

*The SINR coverage probability of the low priority users of synchronous NOMSMT can be expressed as*

$$P_{c,LP}^4 = \sum_{k=1}^{K} \mathcal{A}_k P_{c,LP,k}^4, \tag{55}$$

*where $P_{c,LP,k}^4$ is lower bounded as*

$$P_{c,LP,k}^4 \geq \begin{cases} 0, \ T_{HP} \geq \frac{\alpha_p}{1-\alpha_p}, \\ \frac{2\pi\lambda_k}{\mathcal{A}_k} \int_{r>0} r \exp\left(-\pi \sum_{n=1}^{K} \lambda_n \hat{P}_n^{2/\alpha_2} r^2\right) \\ \times \exp\left(-\frac{(\alpha_2-1)\sigma^2 T_{HP} r^{\alpha_2-1}}{(\alpha_p-(1-\alpha_p)T_{HP})((\alpha_2-1)P_{B_k}+\sum_{n=1}^{K} \lambda_{B_n} P_{B_n} \hat{P}_{B_n}^{(1-\alpha_2)/\alpha_2} r)}\right) dr, \\ \qquad T_{HP} < \frac{\alpha_p}{1-\alpha_p} \ \&\& \ T_{LP} \leq \frac{(1-\alpha_p)T_{HP}}{C}, \\ \frac{2\pi\lambda_k}{\mathcal{A}_k} \int_{r>0} r \exp\left(-\pi \sum_{n=1}^{K} \lambda_n \hat{P}_n^{2/\alpha_2} r^2\right) \\ \times \exp\left(-\frac{(\alpha_2-1)\sigma^2 T_{HP} r^{\alpha_2-1}}{(1-\alpha_p)((\alpha_2-1)P_{B_k}+\sum_{n=1}^{K} \lambda_{B_n} P_{B_n} \hat{P}_{B_n}^{(1-\alpha_2)/\alpha_2} r)}\right) dr, \\ \qquad T_{HP} < \frac{\alpha_p}{1-\alpha_p} \ \&\& \ T_{LP} > \frac{(1-\alpha_p)T_{HP}}{C}. \end{cases} \tag{56}$$

*Proof.* Similar to Theorem 3, replacing $\alpha$ in (44) with $\alpha_1$, (54) can be obtained. The path loss exponent, $\alpha$, in (47) is replaced by $\alpha_2$, to obtain (56) and the proof is completed. □

*3) Average Number of Served Users:*

**Proposition 7.** *The average number of served high and low priority users of synchronous NOMSMT in the typical MBMS area can be expressed, respectively, as*

$$\mathbb{E}^o[N_{HP}^4] = \lambda_B^{-1}\lambda_{U_1}\sum_{k=1}^{K}\mathcal{A}_k P_{c,HP,k}^4, \tag{57}$$

*and*

$$\mathbb{E}^o[N_{LP}^4] = \lambda_B^{-1}\lambda_{U_2}\sum_{k=1}^{K}\mathcal{A}_k P_{c,LP,k}^4. \tag{58}$$

*Proof.* Similar to the asynchronous one, shown in Proposition 3, replacing $P_{c,HP,k}^2$ in (36) and $P_{c,LP,k}^2$ in (37) by $P_{c,HP,k}^4$ and $P_{c,LP,k}^4$, respectively, the corresponding average number of served high and low priority users can be obtained as in (57) and (58). The proof is completed. □

*4) Sum Rate:*

**Proposition 8.** *With fixed rate thresholds for the high and low priority users, $R_{HP}$ and $R_{LP}$, the sum rate for synchronous NOMSMT in the typical MBMS area is*

$$R_{sum}^4 = R_{HP}\lambda_B^{-1}\lambda_{U_1}\sum_{k=1}^{K}\mathcal{A}_k P_{c,HP,k}^4 + R_{LP}\lambda_B^{-1}\lambda_{U_2}\sum_{k=1}^{K}\mathcal{A}_k P_{c,LP,k}^4. \tag{59}$$

*Proof.* Similar to the asynchronous one, shown in Proposition 4, replacing $P_{c,HP,k}^2$ and $P_{c,LP,k}^2$ in (39) by $P_{c,HP,k}^4$ and $P_{c,LP,k}^4$, the sum rate can be obtained as in (59) and the proof is completed.
□

## V. NUMERICAL RESULTS AND DISCUSSIONS

In this section, we will give the numerical results of the SINR coverage probability and sum multicast rate for non-orthogonal MBMS transmission, which are also verified by using Monte Carlo simulations. The related parameters are illustrated in Table I.

Fig. 3 depicts the SINR coverage probabilities of asynchronous and synchronous single-tier networks (STNs), asynchronous and synchronous two-tier HetNets with $\alpha = \{3, 4\}$. The results show that synchronous transmission can significantly improve the SINR coverage probabilities of both single-tier and heterogeneous networks, compared with asynchronous transmission. This is because synchronous transmission enables the BSs to serve the users, instead of contributing the aggregated interference. Moreover, we can observe that the impacts of network density and path loss exponents on synchronous networks are different from those of the asynchronous one. More specifically, in synchronous networks, the SINR coverage probability of the two-tier HetNets is larger than that of STNs, while HetNets cannot improve the SINR coverage probability in the asynchronous one; the SINR coverage probability of path loss exponent, $\alpha = 3$, is larger than that of $\alpha = 4$, while different results are shown in the asynchronous one.

TABLE I
RELATED PARAMETERS OF THE NUMERICAL RESULTS AND SIMULATIONS

| Parameter | Value | Parameter | Value |
|---|---|---|---|
| Carrier frequency | 2 GHz | System bandwidth | 10 MHz |
| Macro BS transmit power | 43 dBm | Small BS transmit power | 30 dBm |
| Macro BS density | $\frac{1}{\pi 1000^2}$ | Small BS density | $\frac{1}{\pi 200^2}$ |
| Path loss exponent | 2.5, 3, 4 | User density | 2000/km$^2$ |

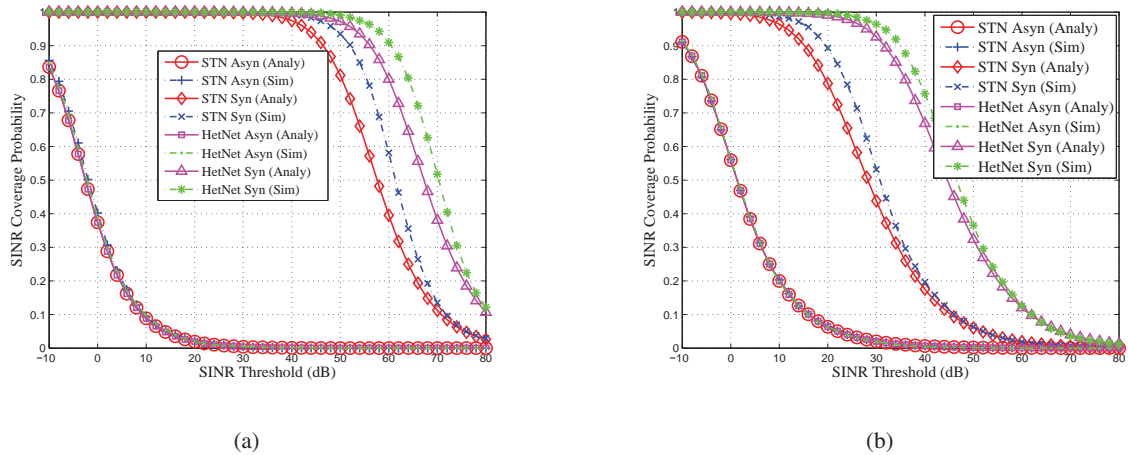

Fig. 3. SINR coverage probabilities of asynchronous and synchronous STNs, asynchronous and synchronous two-tier HetNets: a) $\alpha = 3$; b) $\alpha = 4$

Fig. 4 shows the SINR coverage probability of asynchronous non-orthogonal multi-rate MBMS transmission in a two-tier HetNet with fixed $\alpha = 4$ and different power allocation factors. The results show that in the low SNR threshold region (i.e., the work region of conventional MBMS transmission), non-orthogonal multi-rate MBMS transmission can achieve similar SINR coverage probability as conventional MBMS transmission, as well as provide an extra layer of coverage. However, when the SINR threshold is larger than the maximum SINR detecting the primary layer, $\frac{\alpha_p}{1-\alpha_p}$, the users cannot decode both the primary and secondary layers according to SIC, which causes that the SINR coverage probabilities of both the primary and secondary layers are reduced to zero. The results also show that with more power allocated to the primary layer, the SINR coverage probability of the primary layer is closer to that of conventional MBMS transmission, while that of the secondary layer becomes worse. When all power is allocated to the primary layer, non-orthogonal multi-rate MBMS transmission degrades to conventional one.

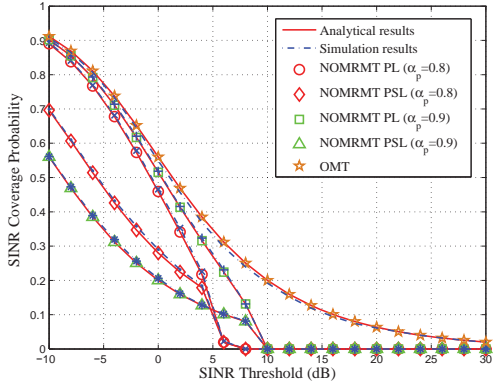
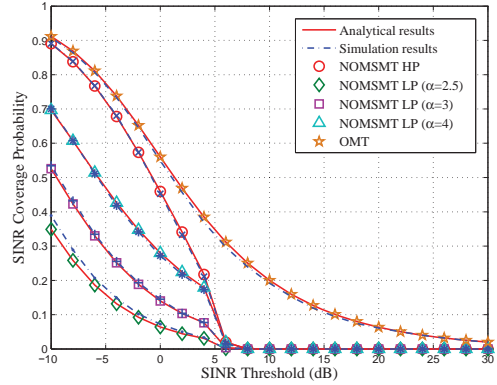

Fig. 4. SINR coverage probability of asynchronous non-orthogonal multi-rate MBMS transmission in a two-tier HetNet

Fig. 5. SINR coverage probability of asynchronous non-orthogonal multi-service MBMS transmission in a two-tier HetNet

With fixed $\alpha_p = 0.8$ and $\alpha = 4$ for conventional MBMS users and high priority users, Fig. 5 illustrates the SINR coverage probability of asynchronous non-orthogonal multi-service MBMS transmission in a two-tier HetNet with different path loss exponents for low priority users. We characterize different types of users, e.g., mobile users and fixed TV users, through assuming different path loss exponents. For example, fixed TV users have smaller path loss exponent than mobile users, due to their roof-top antenna deployment. As can be seen from Fig. 5, the SINR coverage probability of low priority users with path loss exponent, $\alpha = 4$, is best among the given $\alpha = \{2.5, 3, 4\}$, then next is that of $\alpha = 3$, followed by that of $\alpha = 2.5$. This is because with smaller path loss exponent, the users suffer from stronger inter-cell interference in asynchronous networks.

Fig. 6 describes the SINR coverage probabilities of synchronous non-orthogonal multi-rate and multi-service MBMS transmission in a two-tier HetNet, with fixed $\alpha_p = 0.8$, $\alpha = 3$ for the low priority users and $\alpha = 4$ for others. The results show that in the low SINR threshold region, each power layer of both synchronous non-orthogonal multi-rate and multi-service MBMS transmission can achieve the similar SINR coverage probability as synchronous orthogonal one. However, due to power split and the limitation by the maximum SINR detecting the primary layer, $\frac{\alpha_p}{1-\alpha_p}$, with the increase of SINR threshold, the SINR coverage probability of synchronous non-orthogonal MBMS transmission reduces quickly to zero, while the orthogonal one can still provide good network coverage.

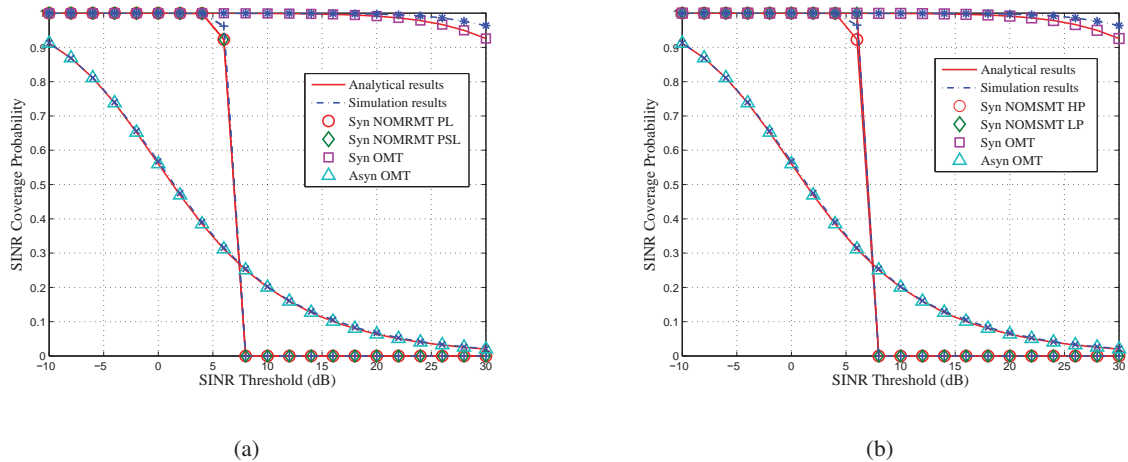

Fig. 6. SINR coverage probabilities of synchronous non-orthogonal MBMS transmission in a two-tier HetNet: a) multi-rate MBMS; b) multi-service MBMS

Fig. 7 plots the sum rate for asynchronous non-orthogonal multi-rate and multi-service conventional MBMS transmission in a two-tier HetNet. With fixed $\alpha_p = 0.8$ and $R_{SL} = \{2, 4\}$ b/s/Hz, the results in Fig. 7(a) show that in the low rate threshold region (i.e., the work region of conventional MBMS transmission), non-orthogonal MBMS transmission can achieve higher sum rate than that of conventional one. This is because non-orthogonal MBMS transmission provides a high data rate secondary layer for the strong users. The results also show that the sum rate for non-orthogonal MBMS transmission with $R_{SL} = 4$ b/s/Hz is smaller than that of $R_{SL} = 2$ b/s/Hz, which reveals that there is an optimal data rate for the secondary layer to achieve the maximum sum rate. For multi-service application, with user densities of high and low priority users, $\lambda_U = \{1000, 1000\}/\text{km}^2$, the results in Fig. 7(b) show that in the low rate threshold region, non-orthogonal MBMS transmission can provide similar service for the high priority users as the conventional one, and also serve the strong users among low priority users simultaneously. With the increase of rate threshold for services, the power domain cannot support two services simultaneously.

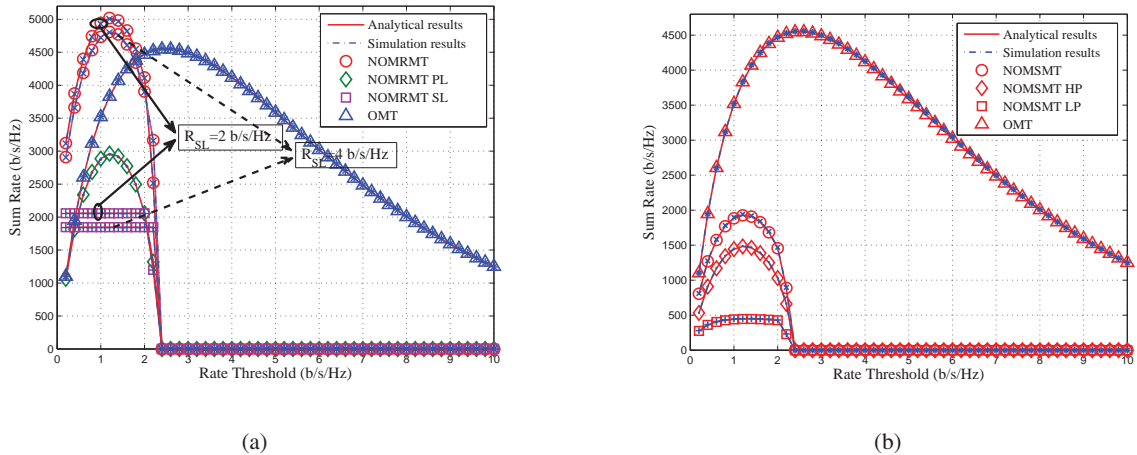

Fig. 7. Sum rate for asynchronous non-orthogonal MBMS transmission in a two-tier HetNet: a) multi-rate MBMS; b) multi-service MBMS

Fig. 8 demonstrates the sum rate for synchronous non-orthogonal multi-rate and multi-service conventional MBMS transmission in a two-tier HetNet. With fixed $\alpha_p = 0.8$ and $R_{SL} = \{2, 4\}$ b/s/Hz, the results in Fig. 8(a) show that in the low rate threshold region, synchronous non-orthogonal multi-rate MBMS transmission can achieve higher sum rate than conventional one. By comparing Fig. 7(a) and Fig. 8(a), synchronous non-orthogonal multi-rate MBMS transmission can achieve higher sum rate than the asynchronous one, and also support higher data rate for the secondary layer. This is because synchronous transmission can significantly improve the quality of the received signal. For multi-service application, with user densities of high and low priority users, $\lambda_U = \{1000, 1000\}/\text{km}^2$, the results in Fig. 8(b) show that synchronous non-orthogonal MBMS transmission can guarantee the QoS of high priority users, and also provide services for low priority users simultaneously. By comparing Fig. 7(b) and Fig. 8(b), synchronous non-orthogonal transmission can provide better coverage for low priority users than the asynchronous one.

## VI. CONCLUSIONS

This paper presented a non-orthogonal MBMS transmission scheme in a $K$-tier single-frequency HetNet, and studied two main usage scenarios: multi-rate and multi-service MBMS transmission. A tractable framework was developed to analyse the performance of non-orthogonal MBMS transmission, by using stochastic geometry. Based on the framework, we derived expressions for the SINR coverage probability, average number of served users, and sum rate. We further considered synchronous transmission to improve the performance of non-orthogonal MBMS delivery, and characterized the different impacts of asynchronous and synchronous transmission on the system performance. Finally, we gave the numerical results verified by Monte Carlo simulation, and drew the main observations from the numerical results as follows: 1) non-orthogonal MBMS transmission can significantly improve the system performance, compared with the conventional one. More specifically, non-orthogonal multi-rate MBMS transmission can fully utilize the difference in channel conditions among users to improve the rate performance, while non-orthogonal multi-service MBMS transmission can efficiently use power resources to guarantee the QoS of the high priority users, and also provide services for the low priority users simultaneously; 2) synchronous non-orthogonal MBMS transmission can achieve superior performance to the asynchronous one, through transforming inter-cell interference into signal transmission. Therefore, non-orthogonal MBMS transmission will be an efficient solution in future wireless networks to satisfying the increasing demands for MBMS. In a future work, it would be of interest to explore the framework for modeling and performance analysis of partially synchronous wireless networks, in which multiple adjacent BSs rather than all BSs perform synchronous MBMS transmission.

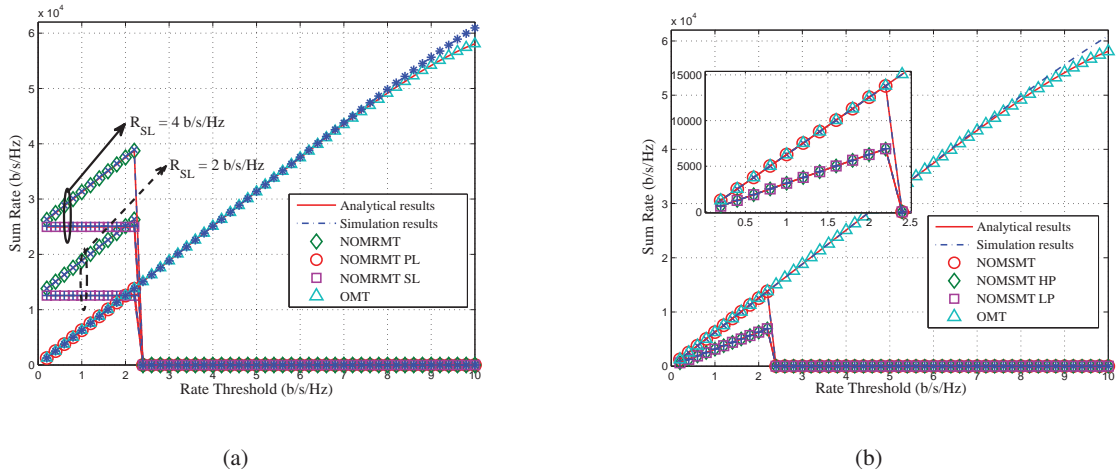

Fig. 8. Sum rate for synchronous non-orthogonal MBMS transmission in a two-tier HetNet: a) multi-rate MBMS; b) multi-service MBMS

## APPENDIX A
## PROOF OF THEOREM 1

For a $K$-tier HetNet, the SINR coverage probability of the primary layer that the typical user is associated with the BS of the $k$-th network tier can be expressed as

$$P_{c,PL,k}^1 = \mathbb{E}_{R_k}[\mathbb{P}[SINR_{PL,k}^1 > T_{PL} \mid R_k = r]] \\ = \int_{r>0} \mathbb{P}[SINR_{PL,k}^1 > T_{PL} \mid R_k = r] f_{R_k}(r) dr, \quad (60)$$

where the probability density function (PDF), $f_{R_k}(r)$, of the distance, $r$, between the typical user and its serving BS is [32]

$$f_{R_k}(r) = \frac{2\pi \lambda_{B_k}}{\mathcal{A}_k} r \exp\left(-\pi \sum_{n=1}^{K} \lambda_{B_n}(\hat{P}_{B_n})^{2/\alpha} r^2\right), \quad (61)$$

where $\mathcal{A}_k$ is the probability that the user is associated with the $k$-th network tier, shown in (14).

Let $C_k = \alpha_{p,k} - (1-\alpha_{p,k})T_{PL}$. Using $SINR_{PL,k}^1$ in (16) and $H_k \sim \exp(1)$, the complementary cumulative distribution function (CCDF) term of $SINR_{PL,k}^1$ can be simplified as

$$\mathbb{P}[SINR_{PL,k}^1 > T_{PL}] = \mathbb{P}\left[\frac{\alpha_{p,k} H_k P_{B_k} r^{-\alpha}}{(1-\alpha_{p,k}) H_k P_{B_k} r^{-\alpha} + \sum_{i=1}^{K} I_{CCI,i} + \sigma^2} > T_{PL}\right]$$

$$= \begin{cases} 0, & T_{PL} \geq \frac{\alpha_{p,k}}{1-\alpha_{p,k}}, \\ \mathbb{E}_{I_{CCI}}\left[\exp\left(-\frac{T_{PL}(\sum_{i=1}^{K} I_{CCI,i} + \sigma^2)}{C_k P_{B_k} r^{-\alpha}}\right)\right], & T_{PL} < \frac{\alpha_{p,k}}{1-\alpha_{p,k}} \end{cases}$$

$$= \begin{cases} 0, & T_{PL} \geq \frac{\alpha_{p,k}}{1-\alpha_{p,k}}, \\ e^{-T_{PL} C_k^{-1} SNR^{-1} r^\alpha} \prod_{i=1}^{K} \mathcal{L}_{I_{CCI,i}}(T_{PL}(C_k P_{B_k})^{-1} r^\alpha), & T_{PL} < \frac{\alpha_{p,k}}{1-\alpha_{p,k}}, \end{cases} \quad (62)$$

where the Laplace transform of $I_{CCI,i}$ is

$$\mathcal{L}_{I_{CCI,i}}(T_{PL}(C_k P_{B_k})^{-1} r^\alpha) = \mathbb{E}_{I_{CCI,i}}\left[\exp(-T_{PL}(C_k P_{B_k})^{-1} r^\alpha I_{CCI,i})\right] \\ = \mathbb{E}_{\Phi_{B_i}}\left[\exp\left(-T_{PL}(C_k P_{B_k})^{-1} \sum_{X_{i,j} \in \Phi_{B_i} \setminus B_{i,o}} H_{i,j} P_{B_i} \|X_{i,j}\|^{-\alpha}\right)\right]. \quad (63)$$

According to [32], the Laplace transform of $I_{CCI,i}$ can be expressed as

$$\mathcal{L}_{I_{CCI,i}}(T_{PL}(C_k P_{B_k})^{-1} r^\alpha) = \exp(-\pi \lambda_{B_i}(C_k^{-1} \hat{P}_{B_i})^{2/\alpha} \mathcal{Z}(T_{PL}, \alpha, C_k) r^2), \quad (64)$$

where

$$\mathcal{Z}(T, \alpha, C_k) = T^{2/\alpha} \int_{(C_k/T)^{2/\alpha}}^{\infty} \frac{1}{1+t^{\alpha/2}} dt. \tag{65}$$

Combining (60), (61), (62), and (64), the SINR coverage probability of the primary layer in the $k$-th network tier can be obtained as

$$P_{c,PL,k}^1 = \begin{cases} 0, \ T_{PL} \geq \frac{\alpha_{p,k}}{1-\alpha_{p,k}}, \\ \frac{2\pi\lambda_{B_k}}{\mathcal{A}_k} \int_0^{\infty} r \exp\left(-T_{PL} C_k^{-1} SNR^{-1} r^{\alpha} \right. \\ \left. -\pi \sum_{i=1}^{K} \lambda_i \hat{P}_{B_i}^{2/\alpha}(1 + C_k^{-2/\alpha}\mathcal{Z}(T_{PL},\alpha,C_k))r^2\right) dr, \ T_{PL} < \frac{\alpha_{p,k}}{1-\alpha_{p,k}}. \end{cases} \tag{66}$$

Combining (18) and (66), the SINR coverage probability of the primary layer can be obtained as in (19).

The SINR coverage probability of the secondary layer that the typical user is associated with the BS of the $k$-th network tier can be expressed as

$$P_{c,PSL,k}^1 = \mathbb{E}_{R_k}[\mathbb{P}[\{SINR_{PL,k}^1 > T_{PL}\} \&\& \{SINR_{SL,k}^1 > T_{SL}\} \mid R_k = r]]$$
$$= \begin{cases} 0, \ T_{PL} \geq \frac{\alpha_p}{1-\alpha_p}, \\ \int_{r>0} \mathbb{P}[SINR_{PL,k}^1 > T_{PL} \mid r] f_{R_k}(r)dr, \ T_{PL} < \frac{\alpha_{p,k}}{1-\alpha_{p,k}} \ \&\& \ T_{SL} \leq \frac{(1-\alpha_{p,k})T_{PL}}{C_k}, \\ \int_{r>0} \mathbb{P}[SINR_{SL,k}^1 > T_{SL} \mid r] f_{R_k}(r)dr, \ T_{PL} < \frac{\alpha_{p,k}}{1-\alpha_{p,k}} \ \&\& \ T_{SL} > \frac{(1-\alpha_{p,k})T_{PL}}{C_k}, \end{cases} \tag{67}$$

where

$$\mathbb{P}[SINR_{SL,k}^1 > T_{SL}] = \mathbb{P}\left[\frac{(1-\alpha_{p,k})H_k P_{B_k} r^{-\alpha}}{\sum_{i=1}^{K} I_{CCI,i} + \sigma^2} > T_{SL}\right]$$
$$= \mathbb{E}_{I_{CCI}}\left[\exp\left(-\frac{T_{PL}(\sum_{i=1}^{K} I_{CCI,i} + \sigma^2)}{(1-\alpha_{p,k})P_{B_k} r^{-\alpha}}\right)\right] \tag{68}$$
$$= e^{-T_{SL}(1-\alpha_{p,k})^{-1}\text{SNR}^{-1}r^{\alpha}} \prod_{i=1}^{K} \mathcal{L}_{I_{CCI,i}}(T_{SL}((1-\alpha_{p,k})P_{B_k})^{-1}r^{\alpha}).$$

Similarly, the Laplace transform of $I_{CCI,i}$ can be expressed as

$$\mathcal{L}_{I_{CCI,i}}(T_{SL}((1-\alpha_{p,k})P_{B_k})^{-1}r^{\alpha}) = \exp(-\pi\lambda_i((1-\alpha_{p,k})^{-1}\hat{P}_{B_i})^{2/\alpha}\mathcal{Z}(T_{PL},\alpha,1-\alpha_{p,k})r^2). \tag{69}$$

Combining (66), (67), (68), and (69), the SINR coverage probability of the secondary layer in the $k$-th network tier can be obtained as

$$P_{c,PSL,k}^1 = \begin{cases} 0, \ T_{PL} \geq \frac{\alpha_{p,k}}{1-\alpha_{p,k}}, \\ \frac{2\pi\lambda_{B_k}}{\mathcal{A}_k} \int_0^\infty r \exp\left(-T_{PL}C_k^{-1}SNR^{-1}r^\alpha \right.\\ \left. -\pi \sum_{i=1}^K \lambda_i \hat{P}_{B_i}^{2/\alpha}(1+C_k^{-2/\alpha}\mathcal{Z}(T_{PL},\alpha,C_k))r^2\right) dr, \\ \qquad T_{PL} < \frac{\alpha_{p,k}}{1-\alpha_{p,k}} \ \&\& \ T_{SL} \leq \frac{(1-\alpha_{p,k})T_{PL}}{C_k}, \\ \frac{2\pi\lambda_{B_k}}{\mathcal{A}_k} \int_0^\infty r \exp\left(-T_{SL}(1-\alpha_{p,k})^{-1}SNR^{-1}r^\alpha \right.\\ \left. -\pi \sum_{i=1}^K \lambda_{B_i} \hat{P}_{B_i}^{2/\alpha}(1+(1-\alpha_{p,k})^{-2/\alpha}\mathcal{Z}(T_{PL},\alpha,1-\alpha_{p,k}))r^2\right) dr, \\ \qquad T_{PL} < \frac{\alpha_{p,k}}{1-\alpha_{p,k}} \ \&\& \ T_{SL} > \frac{(1-\alpha_{p,k})T_{PL}}{C_k}, \end{cases} \quad (70)$$

Combining (18) and (70), the SINR coverage probability of the primary layer can be obtained as in (21).

## APPENDIX B
## PROOF OF PROPOSITION 1

For the $k$-th network tier, the average number of served users by the primary layer in the typical MBMS service area can be expressed as

$$\mathbb{E}^o[N_{PL,k}^1] = \mathbb{E}^o\left[\sum_{y_{k,j}\in\Phi_{U_k,\mathcal{A}(0,R_{TA})}} \mathbb{I}(P_{c,y_{k,j}}(T_{PL}))\right]. \quad (71)$$

According to [6], it can be rewritten as

$$\mathbb{E}^o[N_{PL,k}^1] = \lambda_U \int_{\mathcal{A}(0,R_{TA})} \mathbb{P}[SINR_{PL,k}^1 > T_{PL}]dy = \lambda_B^{-1}\lambda_U P_{c,PL,k}^1. \quad (72)$$

Combining (23) and (72), the average number of served users by the primary layer can be obtained as in (25).

Similarly, the average number of served users by the secondary layer of the $k$-th network tier can be written as

$$\mathbb{E}^o[N_{PSL,k}^1] = \lambda_B^{-1}\lambda_U P_{c,PSL,k}^1, \quad (73)$$

and the average number of served users by the secondary layer can be obtained as in (26). This completes the proof.

## APPENDIX C

## PROOF OF THEOREM 3

To simplify the expressions, let $W = \left| \sum_{k=1}^{K} \sum_{X_{k,j} \in \Phi_{B_k}} P_{B_k}^{1/2} h_{k,j} r_{k,j}^{-\alpha/2} \right|^2$, then $W$ obeys the exponential distribution with mean $\sum_{k=1}^{K} \sum_{X_{k,j} \in \Phi_{B_k}} P_{B_k} r_{k,j}^{-\alpha}$.

For synchronous NOMRMT in a $K$-tier HetNet, the SINR coverage probability of the primary layer that the typical user is associated with the $k$-th network tier can be expressed as

$$P_{c,PL,k}^3 = \mathbb{E}_R[\mathbb{P}[SINR_{PL}^3 > T_{PL}]] = \int_{r>0} \mathbb{P}\left[\frac{\alpha_p W}{(1-\alpha_p)W + \sigma^2} > T_{PL}\right] f_{R_k}(r)dr$$

$$= \begin{cases} 0, \ T_{PL} \geq \frac{\alpha_p}{1-\alpha_p}, \\ \int_{r>0} \mathbb{E}\left[\exp\left(-\frac{\sigma^2 T}{(\alpha_p-(1-\alpha_p))\sum_{n=1}^{K}\sum_{X_{n,j} \in \Phi_{B_n}} P_{B_n} r_{n,j}^{-\alpha}}\right)\right] f_{R_k}(r)dr, \ T_{PL} < \frac{\alpha_p}{1-\alpha_p} \end{cases} \quad (74)$$

$$\geq \begin{cases} 0, \ T_{PL} \geq \frac{\alpha_p}{1-\alpha_p}, \\ \int_{r>0} \exp\left(-\frac{\sigma^2 T}{(\alpha_p-(1-\alpha_p))\mathbb{E}\left[\sum_{n=1}^{K}\sum_{X_{n,j} \in \Phi_{B_n}} P_{B_n} r_{n,j}^{-\alpha}\right]}\right) f_{R_k}(r)dr, \ T_{PL} < \frac{\alpha_p}{1-\alpha_p}, \end{cases}$$

where

$$\mathbb{E}\left[\sum_{n=1}^{K} \sum_{X_{n,j} \in \Phi_{B_n}} P_{B_n} r_{n,j}^{-\alpha}\right] \stackrel{(a)}{=} P_{B_k} r^{-\alpha} + \sum_{n=1}^{K} \int_{x > r\hat{P}_{B_n}^{1/\alpha}} P_{B_n} \lambda_{B_n} x^{-\alpha} dx$$

$$= P_{B_k} r^{-\alpha} + \sum_{n=1}^{K} P_{B_n} \lambda_{B_n} (\alpha-1)^{-1} \hat{P}_{B_n}^{(1-\alpha)/\alpha} r^{1-\alpha}. \quad (75)$$

Note that $(a)$ follows the Campbell's Theorem. Combining (61), (74) and (75), $P_{c,PL,k}^3$ can be obtained as in (44).

Similarly, the SINR coverage probability of the secondary layer that the typical user is associated with the $k$-th network tier can be expressed as

$$P_{c,PSL,k}^3 = \mathbb{E}_R[\mathbb{P}[SINR_{PL}^3 > T_{PL} \ \&\& \ SINR_{SL}^3 > T_{SL}]]$$

$$= \begin{cases} 0, \ T_{PL} \geq \frac{\alpha_p}{1-\alpha_p}, \\ \int_{r>0} \mathbb{P}\left[\frac{\alpha_p W}{(1-\alpha_p)W + \sigma^2} > T_{PL}\right] f_{R_k}(r)dr, \ T_{PL} < \frac{\alpha_p}{1-\alpha_p} \ \&\& \ T_{SL} \leq \frac{(1-\alpha_p)T_{PL}}{C}, \\ \int_{r>0} \mathbb{P}\left[(1-\alpha_p)W > T_{PL}\right] f_{R_k}(r)dr, \ T_{PL} < \frac{\alpha_p}{1-\alpha_p} \ \&\& \ T_{SL} > \frac{(1-\alpha_p)T_{PL}}{C}, \end{cases} \quad (76)$$

where

$$P^3_{c,SL,k} = \int_{r>0} \mathbb{P}\left[(1-\alpha_p)W > T_{PL}\right] f_{R_k}(r)dr$$

$$= \int_{r>0} \mathbb{E}\left[\exp\left(-\frac{\sigma^2 T}{(1-\alpha_p)\sum_{k=1}^{K}\sum_{X_{k,j}\in\Phi_{B_k}} P_{B_k}r_{k,j}^{-\alpha}}\right)\right] f_{R_k}(r)dr. \quad (77)$$

Combining (61), (75) and (77), $P^3_{c,SL,k}$ can be obtained as in (47). The proof is completed.

## REFERENCES

[1] "IMT Vision - Framework and overall objectives of the future development of IMT for 2020 and beyond": ITU-R M.2083-0, Sept. 2015.

[2] J. G. Andrews et al., "What will 5G be?" *IEEE J. Sel. Areas Commun.*, vol. 32, no. 6, pp. 1065-1082, Jun. 2014.

[3] P. Bergmans and T. Cover, "Cooperative broadcasting," *IEEE Trans. Inf. Theory*, vol. 20, no. 3, pp. 317-324, May 1974.

[4] S. R. Mirghaderi, A. Bayesteh, and A. K. Khandani, "On the multicast capacity of the wireless broadcast channel," *IEEE Trans. Info. Theory*, vol. 58, no. 5, pp. 2766-2780, May 2012.

[5] R. O. Afolabi, A. Dadlani, and K. Kim, "Multicast scheduling and resource allocation algorithms for OFDMA-based systems: a survey," *IEEE Commun. Surv. Tuts.*, vol. 15, no. 1, pp. 240-254, 1st Quart. 2013.

[6] X. Lin, R. Ratasuk, A. Ghosh, and J. G. Andrews, "Modeling, analysis, and optimization of multicast device-to-device transmissions," *IEEE Trans. Wireless Commun.*, vol. 13, no. 8, pp. 4346-4359, Apr. 2014.

[7] Z. Wu, V. D. Park, and J. Li, "Enabling device to device broadcast for LTE cellular networks," *IEEE J. Sel. Areas Commun.*, vol. 34, no. 1, pp. 58-70, Jan. 2016.

[8] D. Lecompte and F. Gabin, "Evolved multimedia broadcast/multicast service (eMBMS) in LTE-advanced: overview and Rel-11 enhancements," *IEEE Commun. Mag.*, vol. 50, no. 11, pp. 68-74, Nov. 2012.

[9] A. D. L. Fuente, R. P. Leal, and A. G. Armada, "New technologies and trends for next generation mobile broadcasting services," *IEEE Commun. Mag.*, vol. 54, no. 11, pp. 217-223, Nov. 2016.

[10] L. Dai, B. Wang, Y. Yuan, S. Han, C.-l. I, and Z. Wang, "Non-orthogonal multiple access for 5G: solutions, challenges, opportunities, and future research trends," *IEEE Commun. Mag.*, vol. 53, no. 9, pp. 74-81, Sept. 2015.

[11] Z. Ding, Y. Liu, J. Choi, Q. Sun, M. Elkashlan, C-L. I and H. V. Poor, "Application of non-orthogonal multiple access in LTE and 5G networks," https://arxiv.org/abs/1511.08610.

[12] S. M. R. Islam, N. Avazov, O. A. Dobre, and K.-S. Kwak, "Power-domain non-orthogonal multiple access (NOMA) in 5G systems: potentials and challenges," *IEEE Commun. Surv. Tuts.*, 2016, DOI: 10.1109/COMST.2016.2621116.

[13] Z. Ding, Z. Yang, P. Fan, and H. V. Poor, "On the performance of non-orthogonal multiple access in 5G systems with randomly deployed users," *IEEE Signal Process. Lett.*, vol. 21, no. 12, pp. 1501-1505, Dec. 2014.

[14] Z. Yang, Z. Ding, P. Fan, and G. K. Karagiannidis, "On the performance of non-orthogonal multiple access systems with partial channel information," *IEEE Trans. Commun.*, vol. 64, no. 2, pp. 654-667, Feb. 2016.

[15] P. Xu, Y. Yuan, Z. Ding, X. Dai, and R. Schober, "On the outage performance of non-orthogonal multiple access with one-bit feedback," *IEEE Trans. Wireless Commun.*, vol. 15, no. 10, pp. 6716-6730, Oct. 2016.

[16] Z. Ding, M. Peng, and H. V. Poor, "Cooperative non-orthogonal multiple access in 5G systems," *IEEE Commun. Lett.*, vol. 19, no. 8, pp. 1462-1465, Aug. 2015.

[17] W. Han, Y. Zhang, X. Wang, M. Sheng, J. Li, and X. Ma, "Orthogonal power division multiple access: a green communication perspective," *IEEE J. Sel. Areas Commun.*, 2016, DOI: 10.1109/JSAC.2016.2600139.

[18] Z. Ding, F. Adachi, and H. V. Poor, "The application of MIMO to non-orthogonal multiple access," *IEEE Trans. Wireless Commun.*, vol. 15, no. 1, pp. 537-552, Jan. 2016.

[19] J. Choi, "On the power allocation for MIMO-NOMA systems with layered transmissions," *IEEE Trans. Wireless Commun.*, vol. 15, no. 5, pp. 3226-3237, May 2016.

[20] A. Benjebbour, A. Li, Y. Saito, Y. Kishiyama, A. Harada, and T. Nakamura, "System-level performance of downlink NOMA for future LTE enhancements," in *Proc. IEEE Globecom Workshops (GC Wkshps)*, pp. 66-70, Dec. 2013.

[21] Z. Ding, P. Fan, and H. V. Poor, "Impact of user pairing on 5G non-orthogonal multiple access downlink transmissions," *IEEE Trans. Veh. Technol.*, vol. 65, no. 8, pp. 6010-6023, Aug. 2016.

[22] Z. Yang, Z. Ding, P. Fan, and N. Al-Dhahir, "A general power allocation scheme to guarantee quality of service in downlink and uplink NOMA systems," *IEEE Trans. Wireless Commun.*, vol. 15, no. 11, pp. 7244-7257, Nov. 2016.

[23] B. Di, L. Song, and Y. Li, "Sub-channel assignment, power allocation and user scheduling for non-orthogonal multiple access networks," *IEEE Trans. Wireless Commun.*, vol. 15, no. 11, pp. 7686-7698, Nov. 2016.

[24] J. Choi, "Minimum power multicast beamforming with superposition coding for multiresolution broadcast and application to NOMA systems," *IEEE Trans. Commun.*, vol. 63, no. 3, pp. 791-800, Mar. 2015.

[25] Z. Ding, Z. Zhao, M. Peng, and H. V. Poor, "On the spectral efficiency and security enhancements of NOMA assisted multicast-unicast streaming," https://arxiv.org/abs/1611.05413.

[26] L. Zhang et al., "Layered-division-multiplexing: theory and practice," *IEEE Trans. Broadcast.*, vol. 62, no. 1, pp. 216-232, Mar. 2016.

[27] Q. Zhang, Z. Liang, Q. Li, and J. Qin, "Buffer-aided non-orthogonal multiple access relaying systems in Rayleigh fading channels," *IEEE Trans. Commun.*, DOI: 10.1109/TCOMM.2016.2630050.

[28] Z. Ding, P. Fan, and H. V. Poor, "Random beamforming in millimeter-wave NOMA networks," http://128.84.21.199/abs/1607.06302.

[29] Z. Ding, L. Dai, and H. V. Poor, "MIMO-NOMA design for small packet transmission in the internet of things," *IEEE Access*, vol. 4, pp. 1393-1405, 2016.

[30] Y. Liu, Z. Ding, M. Elkashlan, and J. Yuan, "Non-orthogonal multiple access in large-scale underlay cognitive radio networks," *IEEE Trans. Veh. Technol.*, vol. 65, no. 12, pp. 10152-10157, Dec. 2016.

[31] Y. Liu, Z. Ding, M. Elkashlan, and H. V. Poor, "Cooperative non-orthogonal multiple access with simultaneous wireless information and power transfer," *IEEE J. Sel. Areas Commun.*, vol. 34, no. 4, pp. 938-953, Apr. 2016.

[32] H.-S. Jo, Y. J. Sang, P. Xia, and J. G. Andrews, "Heterogeneous cellular networks with flexible cell association: a comprehensive downlink SINR analysis," *IEEE Trans. Wireless Commun.*, vol. 11, no. 10, pp. 3484-3495, Oct. 2012.

[33] H. S. Dhillon, R. K. Ganti, F. Baccelli, and J. G. Andrews, "Modeling and analysis of $K$-tier downlink heterogeneous cellular networks," *IEEE J. Sel. Areas Commun.*, vol. 30, no. 3, pp. 550-560, Apr. 2012.

[34] J. G. Andrews, F. Baccelli, and R. K. Ganti, "A tractable approach to coverage and rate in cellular networks," *IEEE Trans. Commun.*, vol. 59, no. 11, pp. 3122-3134, Nov. 2011.

[35] J. G. Andrews, A. K. Gupta, H. S. Dhillon, "A primer on cellular network analysis using stochastic geometry," 2016, https://arxiv.org/abs/1604.03183.

[36] J. -R. Ohm, "Advances in scalable video coding," *Proc. IEEE*, vol. 93, no. 1, pp. 42-56, Jan. 2005.